\documentclass[runningheads]{rnl}

\usepackage[utf8]{inputenc}
\usepackage[frenchb]{babel}

\usepackage{graphicx}

\usepackage{amsmath,amsfonts,amssymb}

\begin{document}
 
 \title*{On the dynamics of laminar-turbulent patterns\protect\newline in plane Couette flow}
\titretab{On the dynamics of laminar-turbulent patterns\protect\newline in plane Couette flow}
 \titrecourt{Laminar-turbulent pattern in plane Couette flow}

\author{Paul Manneville}

\index{Manneville Paul}
\auteurcourt{Manneville}

\adresse{Laboratoire d'Hydrodynamique, CNRS UMR 7646, \'Ecole Polytechnique 91128 Palaiseau}

\email{paul.manneville@ladhyx.polytechnique.fr}

\maketitle              

\begin{resume}
L'\'ecoulement de Couette plan pr\'esente des motifs turbulent-laminaire obliques r\'eguliers sur une large plage de nombres de Reynolds $R$ entre l'\'ecoulement de base laminaire globalement stable \`a bas $R<R_{\rm g}$ et le r\'egime uniform\'ement turbulent pour $R>R_{\rm t}$ suffisamment grand.
Les simulations num\'eriques que nous avons pratiqu\'ees sur un motif pr\'esentant une modulation de longueur d'onde montrent une relaxation de cette modulation globalement conforme \`a ce que l'on attend d'une approche standard en termes de structures dissipatives en domaine \'etendu bien que la structuration se d\'eveloppe sur un fond turbulent.
Quelques cons\'equences sont discut\'ees.

\end{resume}

\begin{resumanglais}
Plane Couette flow presents a regular oblique turbulent-laminar pattern over a wide range of Reynolds numbers $R$ between the globally stable base flow profile at low $R<R_{\rm g}$ and a uniformly turbulent regime at sufficiently large $R>R_{\rm t}$.
The numerical simulations that we have performed on a pattern displaying a wavelength modulation show a relaxation of that modulation in agreement with what one would expect from a standard approach in terms of dissipative structures in extended geometry though the structuration develops on a turbulent background.
Some consequences are discussed.
\end{resumanglais}

\section{Introduction\label{S1}}

The diagram below presents the different regimes displayed by plane Couette flow (PCF), the simple shear flow developing between two counter-translating plates at speed $\pm U$ (direction $x$) separated by a gap $2h$ (direction $y$), as a function of the Reynolds number $R=Uh/\nu$, where $\nu$ is the kinematic viscosity of the fluid.
This flow is an archetype of wall-bounded linearly stable flow experiencing a direct transition to turbulence due to finite amplitude perturbations.
In the long term, laminar flow is always recovered for  $R<R_{\rm g}$ but long-lived transients in the form of turbulent spots can be observed down to $R_{\rm u}$.
For $R>R_{\rm t}$ an essentially uniform turbulent regime called {\it featureless\/} is obtained and in the range $R_{\rm g} < R < R_{\rm t}$ an oblique alternation of laminar and turbulent bands is present. Numerical values indicated in the following bifurcation diagram are those of Prigent~\cite{Petal03}:
\begin{figure}[hb!]
\begin{center}
\includegraphics[width=0.65\textwidth ]{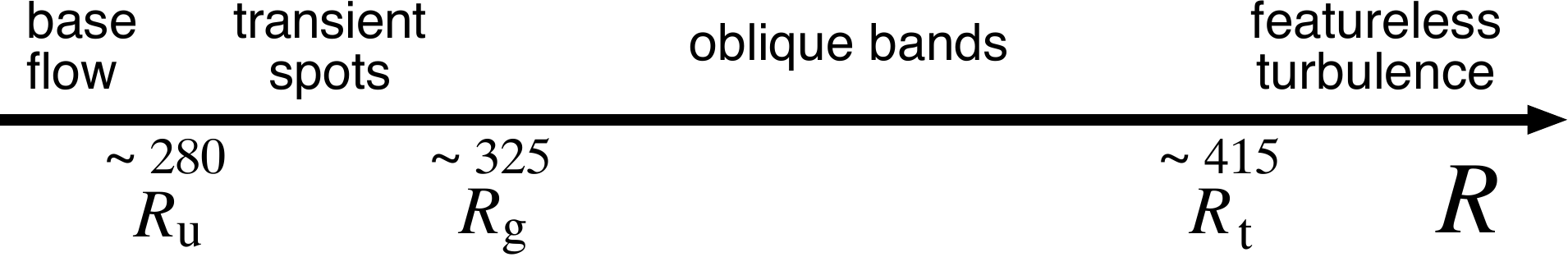}\\
\end{center}
\vspace{-2\baselineskip}
\end{figure}

\noindent Up to now, why bands form as $R$ is decreased below $R_{\rm t}$ is not fully understood despite theoretical efforts~\cite{Tetal09,Ma12} but they are well characterized locally at the statistical level~\cite{BT07} and appropriately described globally at the phenomenological level within the framework of stochastic envelope equations~\cite{Petal03}.
While noise plays a role close to $R_{\rm t}$, its effects become negligible when the bands saturate and the separation of laminar from turbulent local state becomes sharp (see below, Fig.~\ref{F3}).
The aim of this work is to study the relevance of the diffusion formalism to the relaxation of a band pattern presenting an initial phase modulation.
Such a property would be expected from the envelope analysis of patterns in conventional dissipative structure dynamics~\cite{Ma90} when the background over which the pattern develops is laminar and multi-scale analysis can be performed rigorously, at least in principle.
We shall empirically show that it also holds when the patterning develops over a highly fluctuating, turbulent, background.
A similar situation was shown to hold in the barber-pole turbulence regime of transitional cylindrical Couette flow \cite{Hetal89}.
This work is an extension of a previous study focussed on the decay of the band pattern as $R$ is decreased below $R_{\rm g}$~\cite{Ma11}.
The numerical experiment is described in \S\ref{S2} below.
The results are presented in \S\ref{S3} and further discussed in \S\ref{S4}.
 
\section{The experiment\label{S2}}
\begin{figure}[t]  
\begin{center}
\includegraphics[width=0.7\textwidth ]{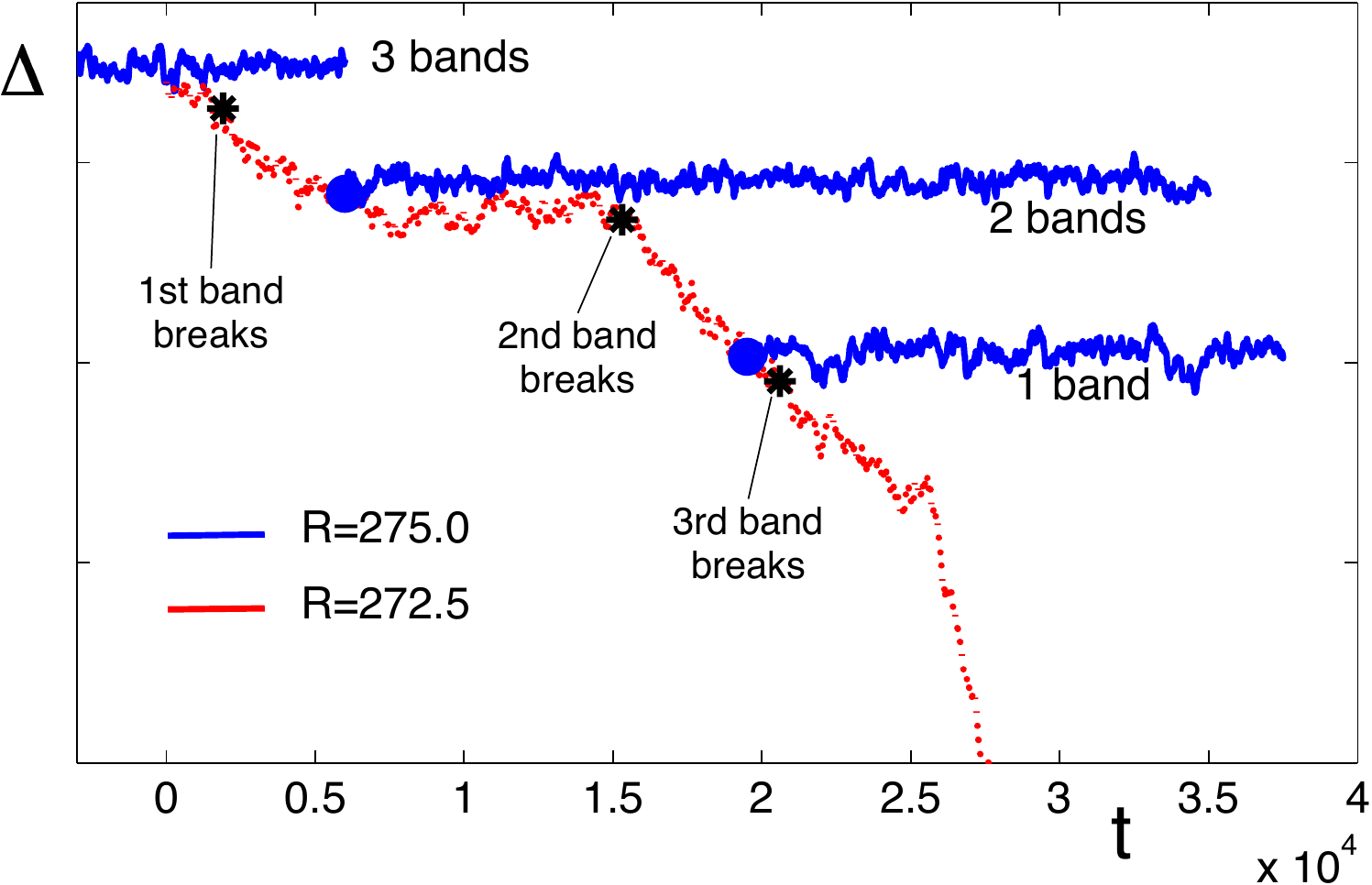}
\end{center}
\caption{Distance to laminar flow $\Delta$ as a function of time during the experiment producing the initial condition. Quantity $\Delta$ is proportional to the kinetic energy content in the perturbation, i.e. the different between the flow field and the reference laminar velocity profile. Consult~\cite{Ma11} and text for details.\label{F2}}
\end{figure}

\begin{figure}[t]
\begin{center}
\begin{minipage}{0.26\textwidth}
\includegraphics[width=\textwidth ]{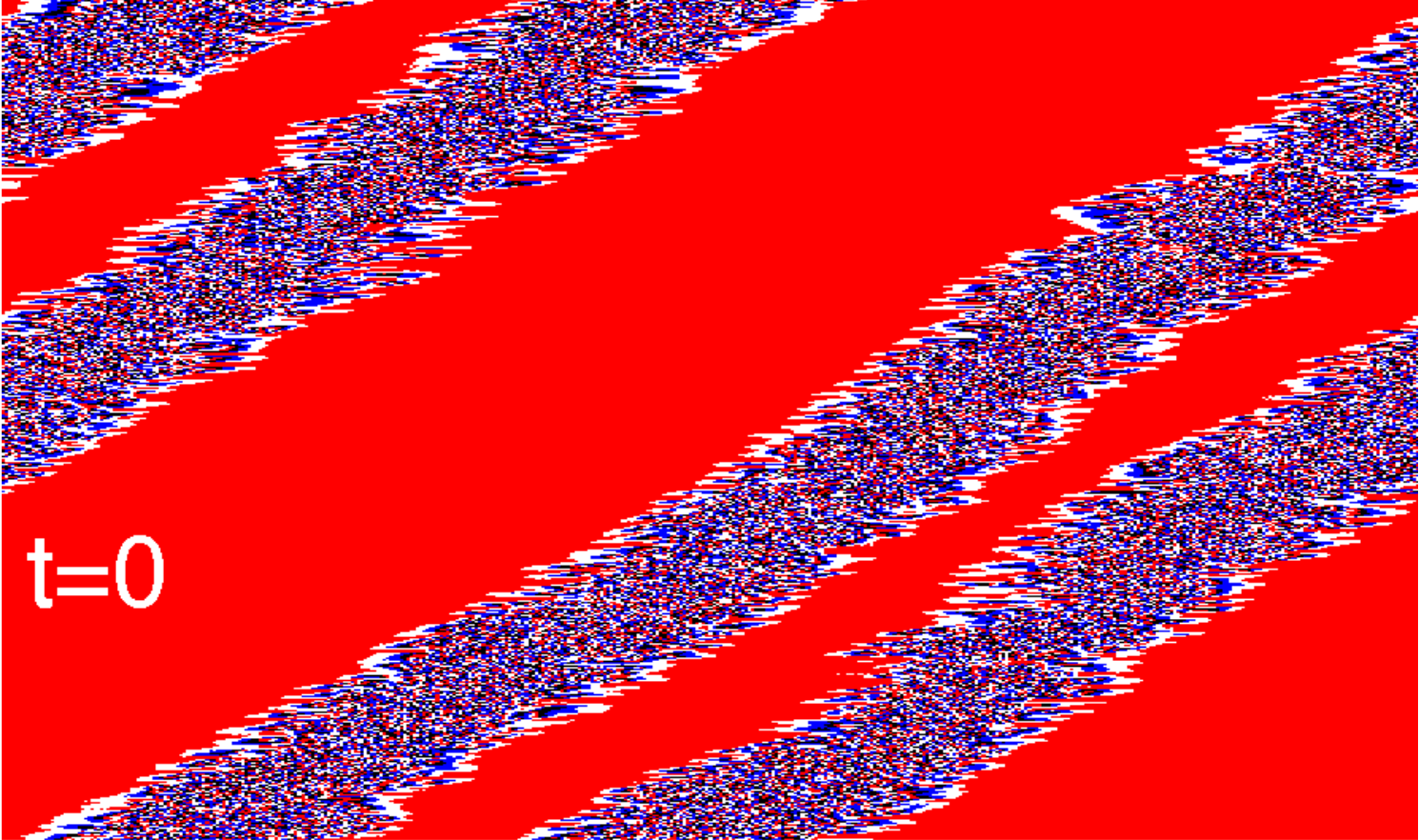}\\[2ex]
\includegraphics[width=\textwidth ]{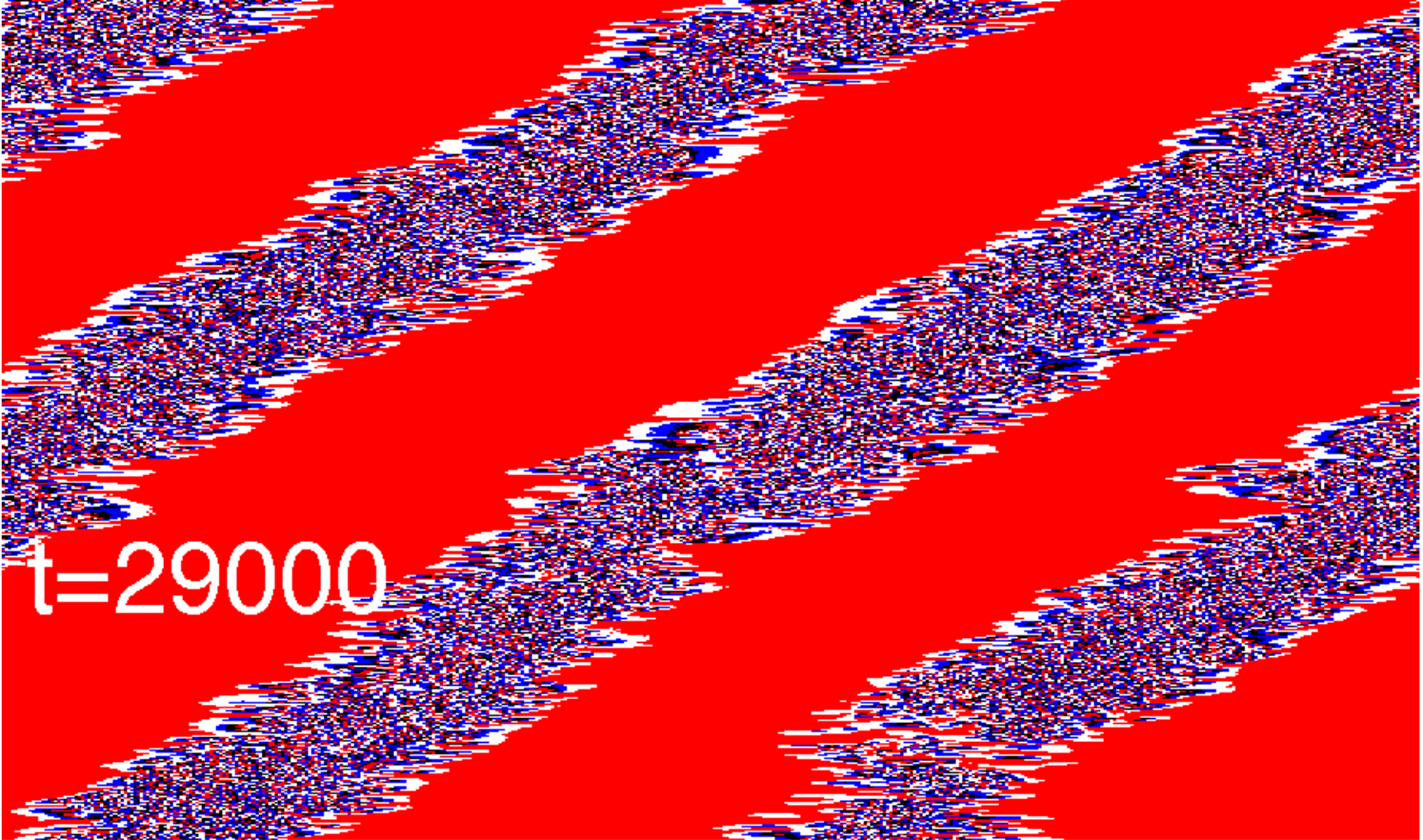}
\end{minipage}
\hspace{0.05\textwidth}
\begin{minipage}{0.5\textwidth}
\includegraphics[width=\textwidth ]{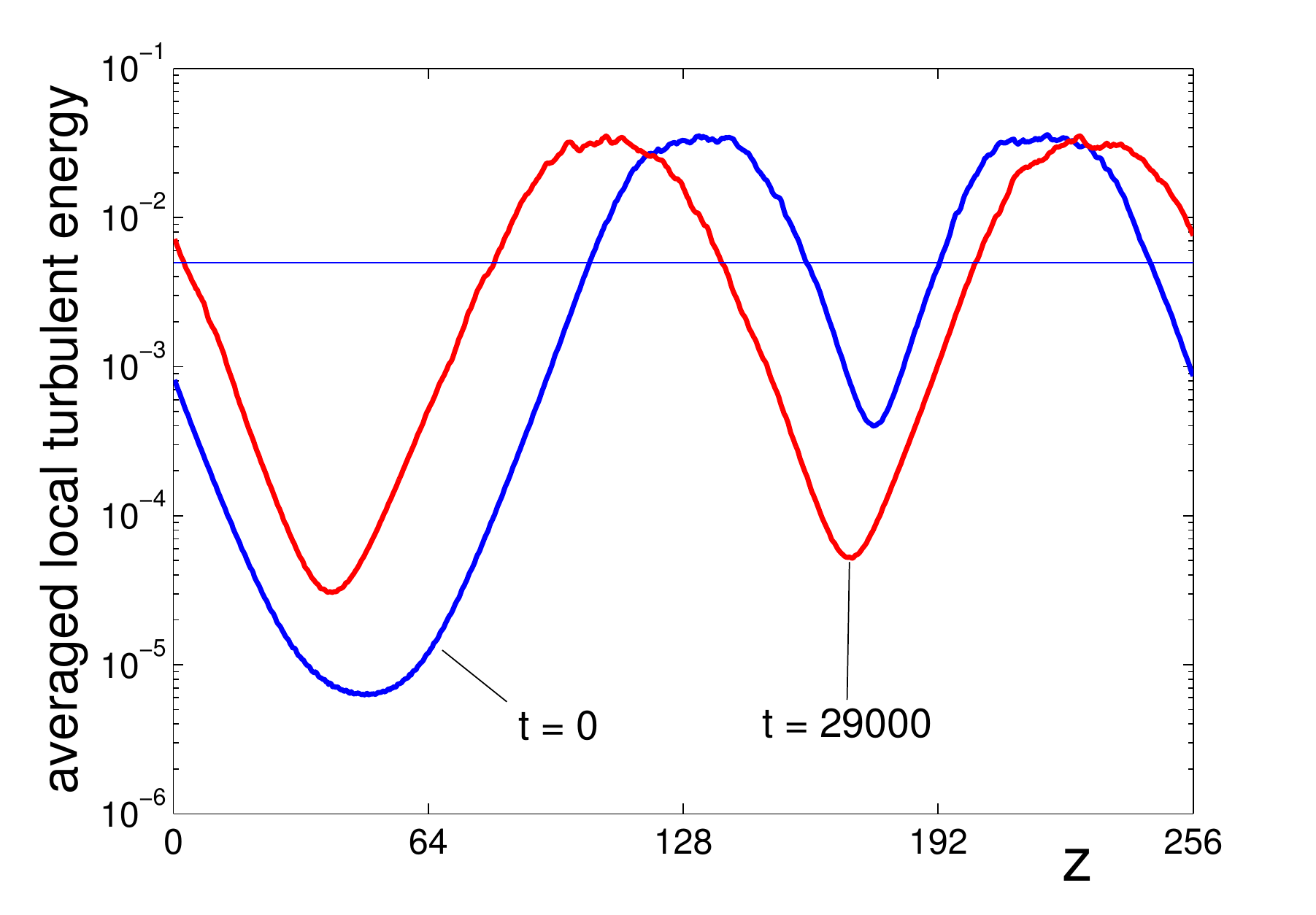}
\end{minipage}
\end{center}
\caption{Snapshots of the bands where the laminar regions appears uniformly in red (left) and turbulent energy profiles averaged along the diagonal of the domain (right) at the start and the end of the experiment. The horizontal line at $0.005$ indicates the cut-off used later to discriminate laminar from turbulent flow.\label{F3}}
\end{figure}

In order to study PCF's turbulent-laminar patterning in conditions relevant to actual laboratory experiments~\cite{Petal03}, one has to consider extended domains, which is computationally demanding.
In order to reduce the numerical load, one can follow Barkley and Tuckerman~\cite{BT07} and consider a narrow but elongated domain aligned with the expected wave-vector of the pattern.
As a drawback, the approach breaks the symmetry of the problem and does not allow orientation fluctuations with respect to the flow's stream-wise direction.
To avoid this limitation, while limiting the computational load, we have considered both stream-wise and span-wise wide domains ($L_x\gg 2h$, $L_z\gg 2h$) with periodic boundary conditions, but at reduced wall-normal resolution.
Previous work has shown that this practice, envisioned as a systematic modeling strategy, leads to a qualitatively satisfactory picture and that the price to be paid is just a downward shift of the transitional range~\cite{MR11}.
To be specific, the experimentally observed $[R_{\rm g},R_{\rm t}]$ interval $\sim [325,415]$ is lowered to $\sim[273.5,360]$ when $N_y=15$ Chebyshev polynomials are used to describe the $y$-dependence of the fields, while it is generally considered that $N_y\ge33$ allows one to fit the experiments back.
Here, we focus on the dynamics of bands as a continuation of the work in~\cite{Ma11} where a domain of size $(L_x\times L_z)=(432 \times 256)$ was numerically simulated using Gibson's code {\sc ChannelFlow} with $N_y=15$. 
For these dimensions and this resolution, a stable pattern of three oblique laminar-turbulent bands is obtained for $R=275$ and, as $R$ is further decreased to $R=272.5$, bands break one after the other, as shown in Fig.~\ref{F2}, where the black asterisks mark the successive breaks.
After a breaking, the remaining turbulent band fragment withdraws regularly, hence a roughly linear decrease of $\Delta$ and, after the fragment has vanished, a pattern with one band less is obtained.
After the first band extinction, if $R=272.5$ is kept, one of the remaining two bands breaks and the process repeats itself (red trace in Fig.~\ref{F2}).
In contrast, taking the states with two or one complete bands obtained during decay as initial conditions (the big blue dots) and increasing back $R$ to $R=275$ allows one to stabilize a 2-band state or a 1-band state besides the original 3-band state (blue traces).
It should be noticed that these subsequent evolutions take place at constant $\Delta$, which mainly means that the {\it amplitude\/} of the pattern is saturated so that its dynamics only involves the position of the bands, the pattern's spatial {\it phase\/}.
We are particularly interested in the 2-band state that displays an irregular band arrangement once the third band has completely receded.
Time is rest to zero at the beginning of the simulation and the initial state is illustrated in Fig.~\ref{F3} (left, top). Its subsequent evolution toward state at $t=29000$ (left, bottom) is scrutinized in the next section.

\section{Results\label{S3}}

\begin{figure}[t!]
\includegraphics[height=0.34\textwidth ]{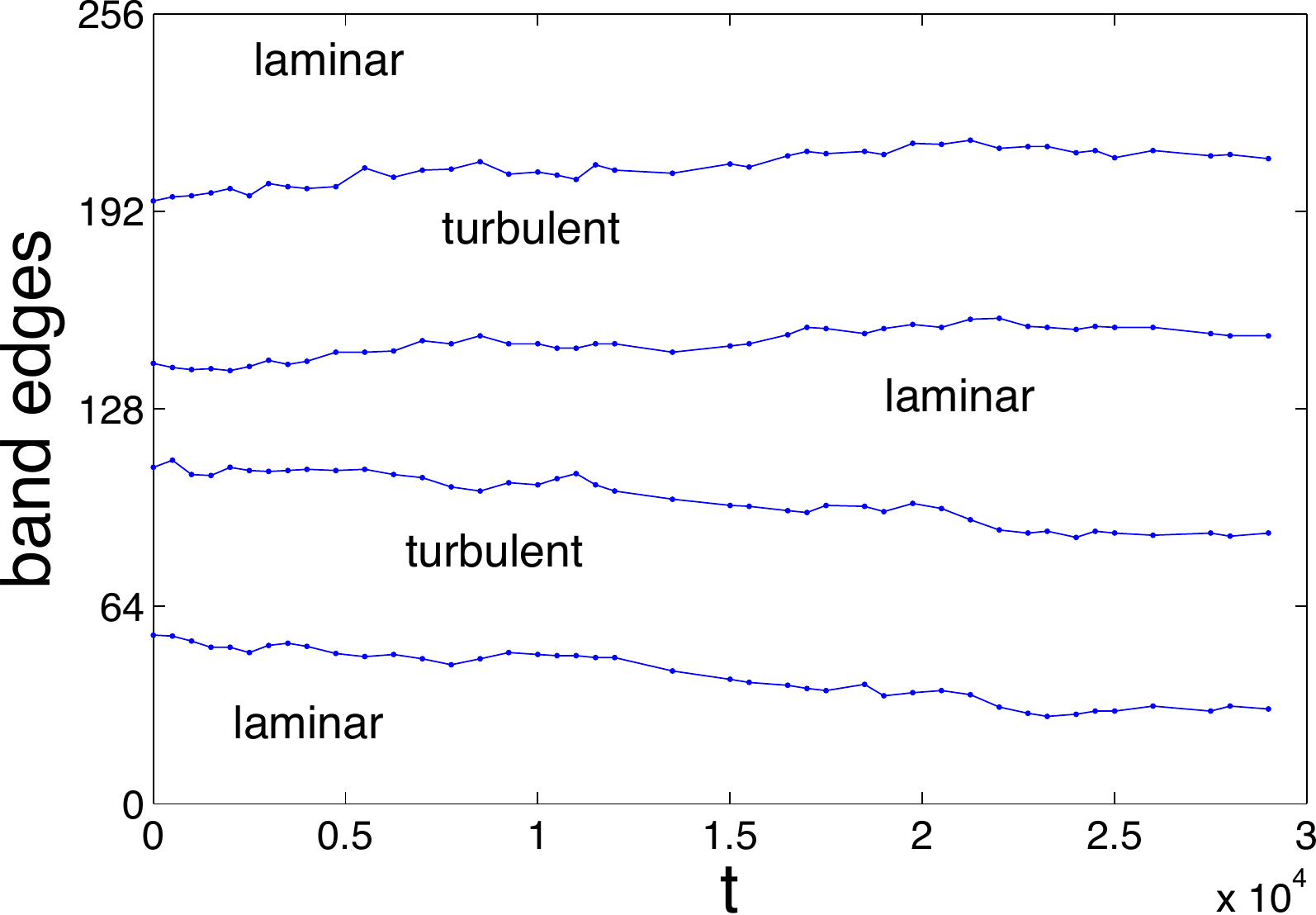}\hfill\includegraphics[height=0.36\textwidth ]{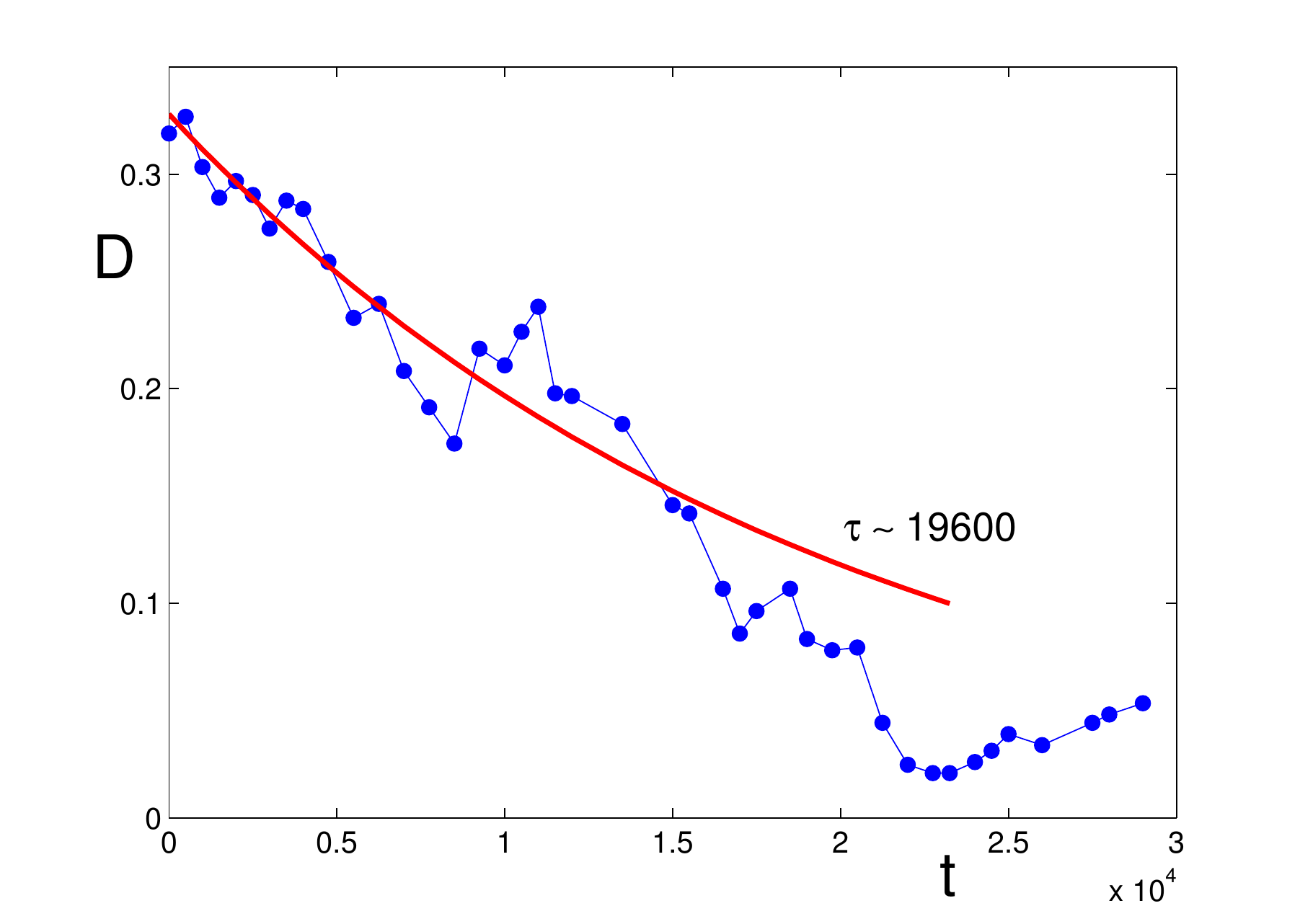}
\caption{Left: Positions of laminar-turbulent interfaces as a function of time during the relaxation of the modulated band pattern at $R=275$. Right: Evolution of the corresponding relative  modulation of the pattern $D$, as defined in the text; the exponential fit is performed on interval $[0,16500]$ yielding a decay time $\tau\sim 19600$.\label{F4}}
\end{figure}

Figure~\ref{F3} (right) displays the mean local perturbation energy $\propto\int {\rm d}y (v_x^2+v_y^2+v_z^2)$, as a function of the span-wise coordinate $z$ after averaging along the bands, thus suppressing dependence on the stream-wise coordinate~$x$. 
It is easily seen that the mean perturbation energy in the turbulent bands is roughly constant all along the evolution and, from the logarithmic scale used, that it decreases exponentially away into the laminar bands.
The band position can thus be unambiguously fixed by choosing a given energy threshold to distinguish laminar from turbulent flow.
Here we place  the cut-off at $0.005$ but results are not sensitive to this value.
Figure~\ref{F4} (left) displays the positions of the laminar-turbulent interfaces as functions of time, showing that the width of the turbulent bands is roughly constant and that they drift so as to equalize the width of the laminar intervals.
We have taken advantage of the periodic boundary conditions to slide the pattern in order to keep it centered, as can be understood from a comparison with Fig.~\ref{F3} (right).
Figure~\ref{F4} (right) displays the relative departure from equalization as a function of time, as measured by quantity $D=(L_z-2\ell)/L_z$, where $\ell$ is the distance between the band centers.
This quantity is indeed seen to decrease to zero as time increases and an exponential behavior can be fitted at the beginning with a decay time constant $\tau \simeq 19600$ but the subsequent  evolution is affected by a high level of noise.

The result cannot be improved by repeating the experiment because a single initial condition is available.
However this state is precisely on the decay trajectory at $R=272.5$ (dotted curve in red in Fig.~\ref{F2}).
On the time interval $\approx[10000,15000]$, after the first broken band has completely disappeared, a phase modulated state nearly identical to that illustrated in Fig.~\ref{F3} (top-left) is obtained.
It again evolves so as  to reduce the modulation up until the second band breaks.
A decay time $\tau\sim 8700$ can be measured, which is of the same order of magnitude as before at $R=275$.
The smaller value can be understood by a larger distance to threshold $R_{\rm t}$, i.e. a more saturated situation with more rigidity and faster relaxation of phase modulations.   
Band breaking is attributed to large deviations towards laminar flow at localized places along the band~\cite{Ma11}.
Provided that the occurrence of such events is sufficiently rare, band drifting should be  observed, which is indeed the case. 

\section{Discussion/Conclusion\label{S4}}

Pattern formation over a turbulent background is an interesting situation developing in transitional plane Couette flow when decreasing the Reynolds number.
In this note we have presented results on a well-formed system of two oblique parallel laminar-turbulent bands with initial wavelength modulation seen to progressively disappear upon evolution, the early stage of which resembles a diffusion process.
Having in mind standard results on patterns described within the envelope formalism, these results are not surprising because they follow from the phase formalism appropriate to deal with cellular instabilities far from threshold, e.g. Rayleigh--B\'enard convection~\cite{Ma90}.
The latter is the prototype of systems producing dissipative structures on a laminar background {\it via\/} a supercritical bifurcation and noise is essentially low level (thermal fluctuations).
Here, if considered at increasing $R$, we face a different situation since the transition to turbulence in PCF is globally subcritical, laminar flow coexisting with turbulent flow all along the transitional range.
This coexistence develops in physical space in the form of an organized pattern and, when considered at decreasing $R$, the emergence of this pattern out of highly fluctuating featureless turbulent flow is continuous and phenomenologically well described by a stochastic  envelope formalism of Langevin--Ginzburg--Landau with tunable external noise mimicking finite and large intrinsic turbulent fluctuations~\cite{Petal03}.
Our preliminary results suggest that processes akin to phase diffusion are similarly at work far below $R_{\rm t}$, close to and above $R_{\rm g}$ where the pattern is well saturated but ready to breakdown.

We now have to formulate reservations that open the field for timely new research.
First, we have a single evidence of wavelength modulation relaxation (two if we count the case at $R=272.5$ interrupted by band breaking).
Next this case cannot be put on precise quantitative footing on the whole available time interval but only at its beginning.
Accordingly, the estimate of the decay constant $\tau$ is not reliable since the fit is performed over a time interval of the same order as $\tau$ itself.
Nevertheless full relaxation is observed, as seen in Fig.~\ref{F4} (right) where $D$ decreases from $\sim 0.3$ down to $\sim 0.05$ for $t> 20000$, which may represent the level of residual fluctuations.

Finally, and most importantly, we must stress that it is obtained through a modeling strategy that consists in reducing the wall-normal resolution in order to be able to simulate large enough domains at reasonable computational cost.
Whereas this strategy is believed to give qualitatively good results~\cite{MR11}, quantitatively it leads to a downward shift of $[R_{\rm g},R_{\rm t}]$, and it is not obvious that it is still effective in describing properties of the long term dynamics of bands, in particular pattern stability and band breaking, at a sufficiently accurate level.
As a matter of fact it was shown in Fig.~\ref{F2} that, within these restrictions, stable bands -- three, two, and even one -- could be observed with $N_y=15$ in a $(432 \times 256)$ periodic domain over time spans much larger than $10^4$ time units ($h/U$).

Recent results  for PCF in a quasi-one-dimensional geometry~\cite{Letal16} show that the decay of turbulence at $R_{\rm g}$ seems well accounted for by a continuous (second order) non-equilibrium phase transition in the directed percolation (DP) class.
Knowing that spatiotemporal correlations are strongly dependent on confinement characteristics~\cite{PM11}, it seems important to scrutinize the quasi-two-dimensional case of interest in experiments.
It seems clear that under-resolved simulations need not be continued and that the indications obtained in this context have to be confirmed by fully resolved computations.
The dynamics of multi-band states and the stability of one-band states while varying the domain size (and its geometrical proportions since, owing to periodic boundary conditions, this controls the angle of bands) should then be studied with two ideas in mind: ({\it i\/}) while decreasing $R$ from $R_{\rm t}$, explore the putative extremum of the potential from which an envelope equation would derive, yielding a preferred wavelength and angle at given $R$, if any;  ({\it ii\/}) examine whether the transition at $R_{\rm g}$ is continuous as implied in the DP scenario, or else discontinuous with finite turbulent fraction at $R_{\rm g}$.
On the one hand, criticality at $R_{\rm g}$ in the continuous case is likely more difficult to ascertain in the quasi-2D case -- ideally requiring a finite-size scaling approach -- than in the quasi-1D case for which convincing results have been obtained~\cite{Letal16}.
On the other hand, a discontinuous transition at $R_{\rm g}$ is forcedly less sensitive to size effects due to the primary role played by extreme localized excursions toward laminar flow~\cite{Ma11}, with a variety of band configurations possible above $R_{\rm g}$, hence a finite range of turbulent fraction depending on the geometry, and no turbulence at all in the long term below $R_{\rm g}$.   
The main merit of our preliminary study at reduced resolution is therefore to help us better characterize the nature of the transition to turbulence and how to describe it at the global level.
\bigskip

\noindent The author would like to thank the members of the JSPS-CNRS collaboration {\sc TransTurb} for discussions related to the work presented here, in particular G. Kawahara and M. Shimizu (Osaka), T. Tsukahara (Tokyo) on the Japanese side, and Y. Duguet (LIMSI), R. Monchaux and M. Couliou (ENSTA-ParisTech) on the French side.


\begin{thebibliography}{let1}

\bibitem{BT07}
{\sc D. Barkley}, {\sc L.S. Tuckerman}
Mean flow of turbulent-laminar patterns in plane Couette flow, 
{\it J. Fluids Mech.}, {\bf 576}, 109--137 (2007).

\bibitem{Hetal89}
{\sc J.J. Hegseth}, {\sc C.D. Andereck}, {\sc F. Hayot}, {\sc Y. Pomeau},
Spiral turbulence and phase dynamics
{\it Phys. Rev. Lett.} {\bf 62} (1989) 257--260.

\bibitem{Letal16}
{\sc G. Lemoult}, {\sc L. Shi}, {\sc K. Avila}, {\sc S.V. Jalikop}, {\sc M. Avila}, {\sc B. Hof},
Directed percolation phase transition to sustained turbulence in Couette flow.
{\it Nature Physics\/}, to appear (2016).

\bibitem{Ma90}
{\sc P. Manneville}, {\it Dissipative structures and weak turbulence\/}, Academic Press (1990).

\bibitem{MR11}
{\sc P. Manneville}, {\sc J. Rolland}, 
On modelling transitional turbulent flows using under-resolved direct numerical simulations: The case of plane Couette flow,
{\it Theor. Comput. Fluid Dyn.} {\bf 25} (2011) 407--420.

\bibitem{Ma11} 
{\sc P. Manneville},
On the decay of turbulence in plane Couette flow, 
{\it Fluid Dyn. Res}, {\bf 43}, 065501 (2011).

\bibitem{Ma12}
{\sc P. Manneville},
Turbulent patterns in wall-bounded flows: a Turing instability?
{\it EPL\/} {\bf 98}, 64001 (2012).

\bibitem{PM11}
{\sc J. Philip}, {\sc P. Manneville},
From temporal to spatiotemporal dynamics in transitional plane Couette flow,
{\it Phys. Rev. E} {\bf 83}, 036308 (2011).

\bibitem{Petal03}
{\sc A. Prigent}, {\sc G. Gregoire}, {\sc H. Chat\'e}, {\sc O. Dauchot},
Long wavelength modulation of turbulent shear flows,
{\it Physica D}, {\bf174}, 100--113 (2003).

\bibitem{Tetal09}
{\sc L.S. Tuckerman}, {\sc D. Barkley}, {\sc O. Dauchot},
Instability of uniform turbulent plane Couette flow: spectra, probability distribution functions and $K-\Omega$ closure model,
{\it 7th IUTAM Symposium on Laminar-Turbulent Transition}, Stockholm
Ph. Schlatter, D. Henningson, eds. Springer (2009) pp. 59-66.

\end{thebibliography}
\end{document}